\documentclass[pra]{revtex4}
\usepackage{eurosym}
\usepackage{amssymb}
\usepackage{amsmath}
\usepackage{graphicx}
\usepackage{dcolumn}
\usepackage[center]{subfigure}
\usepackage{color}

\begin{document}

\title{Spatiotemporal solitary modes in a twisted cylinder waveguide pipe with the self-focusing Kerr nonlinearity}

\author{Hao Huang$^{1}$, Lin Lyu$^{1}$, Maobin Xie$^{2}$, Weiwen Luo$^{2}$, Zhaopin Chen$^{3}$, Zhihuan Luo$^{2}$, Chunqing Huang$^{1}$, Shenhe Fu$^{4}$, and Yongyao Li$^{1}$}
\email{yongyaoli@gmail.com}
\affiliation{$^{1}$School of Physics and Optoelectronic Engineering, Foshan University,
Foshan $528000$, China\\
$^{2}$Department of Applied Physics, South China Agricultural University, Guangzhou 510642, China\\
$^{3}$Department of Physical Electronics, School of Electrical Engineering,
Faculty of Engineering, and the Center for Light-Matter Interaction, Tel
Aviv University, Tel Aviv 69978, Israel\\
$^{4}$Department of Optoelectronic Engineering, Jinan University, Guangzhou 510632, China}




\begin{abstract}
We study the spatiotemporal solitary modes that propagate in a hollow twisted cylinder waveguide pipe with a self-focusing Kerr nonlinearity. Three generic solitary modes, one belonging to the zero-harmonic (0H) and the other two belonging to the first-harmonic (1H), are found in the first rotational Brillouin zone. The 0H solitary modes can be termed as a quasi-1D (one-dimensional) temporal soliton. Their characteristics depend only on the energy flow. The 1H solitary mode can be termed a quasi-2D (two-dimensional) bullet, whose width is much narrower than the angular domain of the waveguide. In contrast to the 0H mode, the characteristics of the 1H solitary mode depend on both their energy flow and the rotating speed of the waveguide. We demonstrate numerically that the 1H solitary modes are stable when their energy flow is smaller than the threshold norm of the \emph{Townes soliton}. The boundaries of the bistable area for these two types of solitary modes are predicted by the analyses via two-mode approximation. This prediction is in accordance with the numerical findings. We also demonstrate analytically that the 1H solitary mode of this system can be applied to emulate the nonlinear dynamics of solitary modes with 1D Rashba spin-orbit (SO) coupling by optics. Two degenerated states of the 1H solitary mode, semi-dipole and mixed mode, are found from this setting via the mechanism of SO coupling. Collisions between the pair of these two types of solitary modes are also discussed in the paper. The pair of the 0H solitary mode features only the elastic collision, whereas the pair of 1H solitary modes can feature both elastic and inelastic collision when the total energy flow of the two modes are smaller or close to the threshold norm of the \emph{Townes soliton}.
\end{abstract}
\maketitle


\section{Introduction}

Producing a two-dimensional (2D) solitary wave in free pace has attracted great interest in many branches of nonlinear sciences. In optics, 2D solitary waves can be 2D spatial solitons propagating in a bulk medium or spatial temporal solitons, e.g., light bullets, propagating in the planar waveguides. There is a commonly known difficulty: the fundamental 2D solitary mode in homogeneous media with ubiquitous cubic self-focusing Kerr (cubic self-attractive) nonlinearity is destabilized by the critical collapse \cite{Berg1998}. To overcome this difficulty, diverse methods have been proposed to stabilize 2D solitary modes. These methods are mainly based on replacing the cubic attractive nonlinearity with another non-cubic type nonlinearities that do not give rise to the collapse.
 In particular, stable 2D optical solitons have been predicted and created in media with saturable \cite{Segev1994}, quadratic \cite{Mihalache20062}, cubic-quintic \cite{Mihalache2006} and nonlocal nonlinearities \cite{Mihalache20063,Skupin2006}. However, for homogeneous media with the pure self-focusing Kerr effect, determining how to stabilize a 2D optical soliton remains a challenging issue.

Recently, an unexpected result was reported: 2D matter-wave solitons can be stabilized by the Rashba spin-orbit (SO) coupling for the binary Bose-Einstein condensates with a cubic self-attractive nonlinearity \cite{Sakaguichi2014,Sakaguichi2016}. Based on this finding, a similar mechanism for an optical system was applied to stabilize a spatiotemporal soliton in media with the self-focusing Kerr nonlinearity \cite{Kartashov2015,Kartashov20152}. This system is realized by the planar dual-core nonlinear waveguide with temporal dispersive coupling between the cores. Note that the temporal dispersive coupling in this setting is equivalent to the 1D (one-dimensional) SO coupling of the Dresselhaur type \cite{Zhaihui2015}. Another similar mechanism for optical system, which is based on the coupled planar waveguide with the diffraction coupling, also supports a 2D spatiotemporal soliton in media with self-focusing Kerr nonlinearity \cite{SakaguichiNJP2016}. The diffraction coupling in this system is equivalent to the 1D SO coupling of the Rashba type \cite{Yongping,Su2016}. From these works, one can find that the differential coupling, which is an element of the SO coupling, may help stabilize 2D optical solitons in self-focusing Kerr media. However, a necessary condition for these works is that they must be realized on two-component systems; otherwise, such couplings appear impossible to create. An interesting consideration, which seems unreasonable, is the realization of such coupling in a single-component optical system.

\begin{figure}
\centering  \includegraphics[scale=0.6]{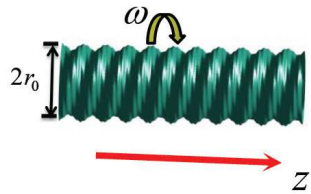}
\caption{(Color online) Sketch of the twisted cylinder waveguide shell.}
\label{sketch}
\end{figure}

The objective of this work is to create SO-coupling in a single component system, which can be realized on a twisted hollow waveguide pipe with a screw pitch $2\pi/\omega$ (see Fig. \ref{sketch}) and a self-focusing Kerr nonlinearity. In fact, this twisted system is a type of rotational nonlinear system. It is well known that rotational nonlinear systems feature many specific nonlinear dynamics in terms of both optics and Bose-Einstein condensates \cite{Chiao1986,Tomita1986,Kopp2003,Ornigotti2007,Kanamoto2008,Bhat2006,Reijnders2004,Pu2005,Kasamatsu2006,Saito2004,Schwartz2006,Wen2012,Wen2010,Wilkin2000,Cooper2001}. By applying coordinate transformation, the dynamics of the corresponding optical field can be transferred from the quiescent reference frame to the rotational reference frame, and a differential operator is created in the nonlinear Schr\"{o}dinger equation by this transformation. Although this differential term cannot create a differential coupling by itself, we demonstrate that a Rashba type SO coupling, occurring between the real part and the imaginary part of the complex mode, can be created by introducing the periodical boundary condition, which is automatically satisfied for this system. We find that a stable quasi-2D solitary mode does exist under this circumstance. To clarify this mechanism, another objective of this work is to identify all of the solitary modes that exist in the first rotational Brillouin zone (FRBZ) \cite{Li2012,Guihua2013,Yongyao2014,Haoxu2014,Guihua2017,Pangwei2014}, from which all solitary modes are boosted out.
 Three generic solitary modes are found in this zone: one belongs to the zero-harmonic (0H), and the other two belong to the first-harmonic (1H). Two 1H solitary modes are semi-dipoles and mixed modes, which can be characterized by the mechanism of the Rashba SO coupling.  The characteristics of these solitary modes are discussed systematically through the paper. Our paper is structured as follows: the models and the theoretical analysis of the solitary modes are illustrated in Sec. II. A numerical simulation and analysis to the solutions of two types of solitary modes are studied in Sec. III. The collisions between a pair of solitary modes are discussed in Sec. IV, and the paper is concluded in Sec. V.

\section{Model and the analysis}

The sketch map of our setting is displayed in Fig. \ref{sketch}: a hollow cylinder waveguide pipe that features a self-focusing Kerr nonlinearity is twisted with a screw pitch of $2\pi/\omega$. If the width of pipe shell is sufficiently thin, the variable $r$ can be fixed by the radius of the pipe, $r_0$, which can be normalized to $r_{0} = 1$. Hence, the 2D Laplace operator in the polar coordinates form is reduced to $\nabla^{2}\sim(1/2\partial^{2}/\partial\theta^{\prime2})$ (where $\theta^{\prime}$ is the stationary angular coordinate). The scaled form of dynamics of the spatial-temporal nonlinear wave propagating such optical setting is governed by the 2D nonlinear Schr\"{o}dinger equation, in the rotating reference frame (or in the helical coordinate system) as
\begin{eqnarray}
i{\partial u\over\partial z}=-{1\over2}\left({\partial^{2}u\over\partial\theta^{2}}+{\partial^{2}u\over\partial T^{2}}\right)+i\omega{\partial u\over\partial\theta}-|u|^{2}u\label{NLS}
\end{eqnarray}
where $\theta=\theta^{\prime}-\omega z$ is the helical coordinate and $\omega$ is the rotation speed of the system. Stationary 2D solitary modes of this  system with the propagation constant $K$ is sought as $u(z,\theta,T)=e^{i K z}\phi(\theta,T)$, with the function $\phi(\theta,T)$ obeying
\begin{eqnarray}
-{1\over2}\left({\partial^{2}\phi\over\partial\theta^{2}}+{\partial^{2}\phi\over\partial T^{2}}\right)+i\omega{\partial \phi\over\partial\theta}-|\phi|^{2}\phi+K\phi=0.\label{NLS2}
\end{eqnarray}
Equation (\ref{NLS}) conserves the energy flow of the solitary modes:
\begin{eqnarray}
P=\int^{\infty}_{-\infty}dT\int^{\pi}_{-\pi}d\theta|u(\theta,T)|^{2}.\label{Power}
\end{eqnarray}

The stability of the solitary mode can be identified by inputting the perturbed solution from
\begin{eqnarray}
u(\theta,T,z)=e^{i K z}\left[\phi(\theta,T)+v_{1}(\theta,T)e^{-i\lambda z}+v_{2}^{\ast}(\theta, T)e^{i\lambda^{\ast} z }\right],\label{perturbation}
\end{eqnarray}
where $v_{1,2}$ are perturbation eigenmodes, $\lambda$ is the corresponding eigenvalue and the asterisk denotes the complex conjugate. The substitution of expression (\ref{perturbation}) into Eq. (\ref{NLS}) and linearization leads to the eigenvalue problem perturbation frequency, $\lambda=\lambda_{\mathrm{r}}+i\lambda_{\mathrm{i}}$, and the eigenfunctions, $\{v_{1},v_{2}\}$:

\begin{equation}
\left(\begin{array}{cc}
A_{11} & A_{12} \\
A_{21} & A_{22}
\end{array}\right)
\left(
\begin{array}{c}
v_{1}\\
v_{2}
\end{array}\right)=\lambda\left(
\begin{array}{c}
v_{1}\\
v_{2}
\end{array}
\right),\label{eigenvalue}
\end{equation}
where $A_{11}=-(1/2)(\partial_{\theta\theta}+\partial_{TT})+i\omega\partial_{\theta}+K-2|\phi|^{2}$, $A_{12}=-\phi^{2}$, $A_{21}=-A^{\ast}_{12}$ and $A_{22}=-A^{\ast}_{11}$. The underlying solution $\phi$ is stable if all the eigenvalues are real.

According to the analyses in Ref. \cite{Li2012}, Eqs. (\ref{NLS},\ref{NLS2}) are invariant with respect to the boost transformation when applying the periodic boundary condition $u(\theta,T)=u(\theta+2n\pi,T)$ (where $n$ is arbitrary integer). Such invariance allows one to change the rotation speed from $\omega$ to $\omega-n$. Therefore, the rotational speed can be restricted to the interval $\omega\in[0,1]$. Furthermore, by applying an additional invariance: $u(\theta,T,z;\omega)=u^{\ast}(\theta,T,-z;-\omega)$, which is also admitted by Eqs. (\ref{NLS},\ref{NLS2}), the rotation speed may be eventually restricted to a narrower interval,
\begin{eqnarray}
0\leq\omega \leq 1/2, \label{FRBZ}
\end{eqnarray}
which was called the first rotational Brillouin zone (FRBZ)\cite{Li2012,Guihua2013,Yongyao2014,Haoxu2014,Guihua2017,Pangwei2014}.

To study the transverse mode in this system in domain of $\theta$, we drop the dependence of $T$ (i.e., assume that the field is homogeneous on $T$) and the nonlinearity (i.e., consider it in the $0$-order case) first from Eq. (\ref{NLS2}); therefore, equation (\ref{NLS2}) changes to
\begin{eqnarray}
-K\phi=-{1\over2}\phi''+i\omega\phi'.\label{0order}
\end{eqnarray}
In fact, equation (\ref{0order}) is the linear limit of the Eq. (\ref{NLS2}). Assume the field can be defined by $\phi=e^{in\theta}$, where $n=0,1,2,\ldots$, which refers to, e.g.,  the zero harmonic (0H), first harmonic (1H), and second harmonic (2H) mode, and substituting it into Eq. (\ref{0order}), one obtains
\begin{eqnarray}
-K={1\over2}n^{2}-\omega n.\label{harmoniceq}
\end{eqnarray}
Assume two different harmonic modes $n_{1}$ and $n_{2}$ ($n_{1}\neq n_{2}$) are bistable in the FRBZ, which may satisfy
\begin{eqnarray}
{1\over2}n^{2}_{1}-\omega n_{1}={1\over2}n^{2}_{2}-\omega n_{2}. \label{bistableharmonic}
\end{eqnarray}
One may obtain $\omega=(n_{1}+n_{2})/2$ from Eq. (\ref{bistableharmonic}). In the interval of $\omega\in[0,0.5]$, only $\omega=0.5$ satisfies Eq. (\ref{bistableharmonic}) for $n_{1}=0$ and $n_{2}=1$. This analysis predicts that only the 0H and 1H modes can be found in the FRBZ in the linear limit. In the next section, the numerical simulations demonstrate that 0H and 1H spatial-temporal solitary modes are found in this system. The 0H solitary mode exists in the full range of the FRBZ with a homogeneous amplitude in the domain of $\theta$; therefore, the solitary mode of this type can be termed a quasi-1D temporal soliton. The 1H solitary mode bifurcates from the right edge of FRBZ, i.e., $\omega=0.5$ when $P\rightarrow0$; they are complex functions and have a single-peak structure both in the $\theta$ and $T$ domain. Therefore, the solitary mode of this type can be termed a quasi-2D bullet.

\section{Numerical simulation and theoretical analysis of the 0H and 1H temporal soliton in the system}
\begin{figure}
\centering  \includegraphics[scale=0.5]{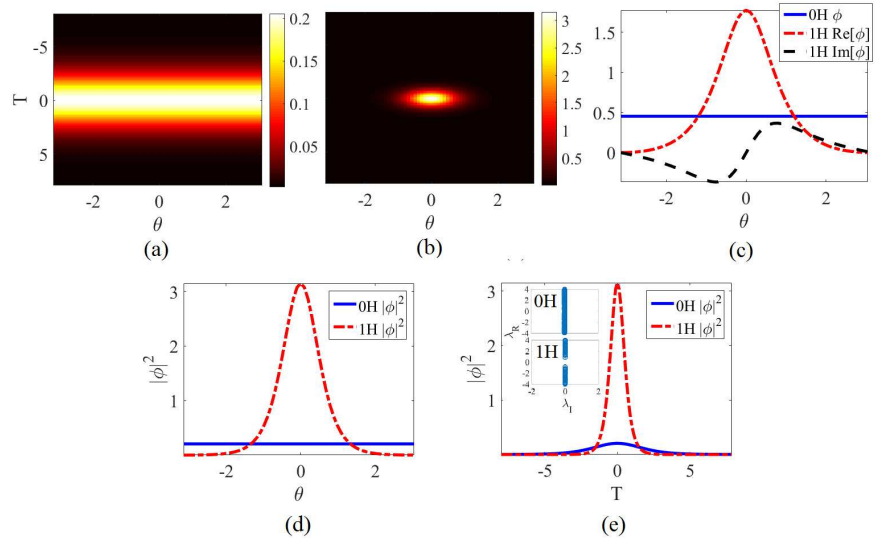}
\caption{(Color online) (a) A typical stable 0H solitary mode (quasi-1D temporal soliton) with $(P,\omega)=(5.8,0.5)$. (b) A typical stable 1H solitary mode (quasi-2D bullet) with the same magnitude of $(P,\omega)$ of panel (a). (c) Cross-sections of the amplitudes of 0H mode (the bule solid line), which is real and a constant in the $\theta$ domain, and the amplitude of the 1H mode in the same domain, which is complex (the red dash-dot curve is the real part, and the black dashed curve is the imaginary part). (d) Cross-sections of the intensity profile of the 0H (blue solid line) and 1H (red dashed curve) modes in the $\theta$ domain. (e) Cross-sections of the intensity profile of the 0H (blue solid curve) and 1H (red dashed curve) modes in the $T$ domain. The two insets in panel (e) are the spectra of the eigenvalues of these two types of solitary modes (upper panel for the 0H mode and lower panel for the 1H mode), which demonstrate their stabilities.}
\label{example}
\end{figure}
\begin{figure}
\centering  \includegraphics[scale=0.15]{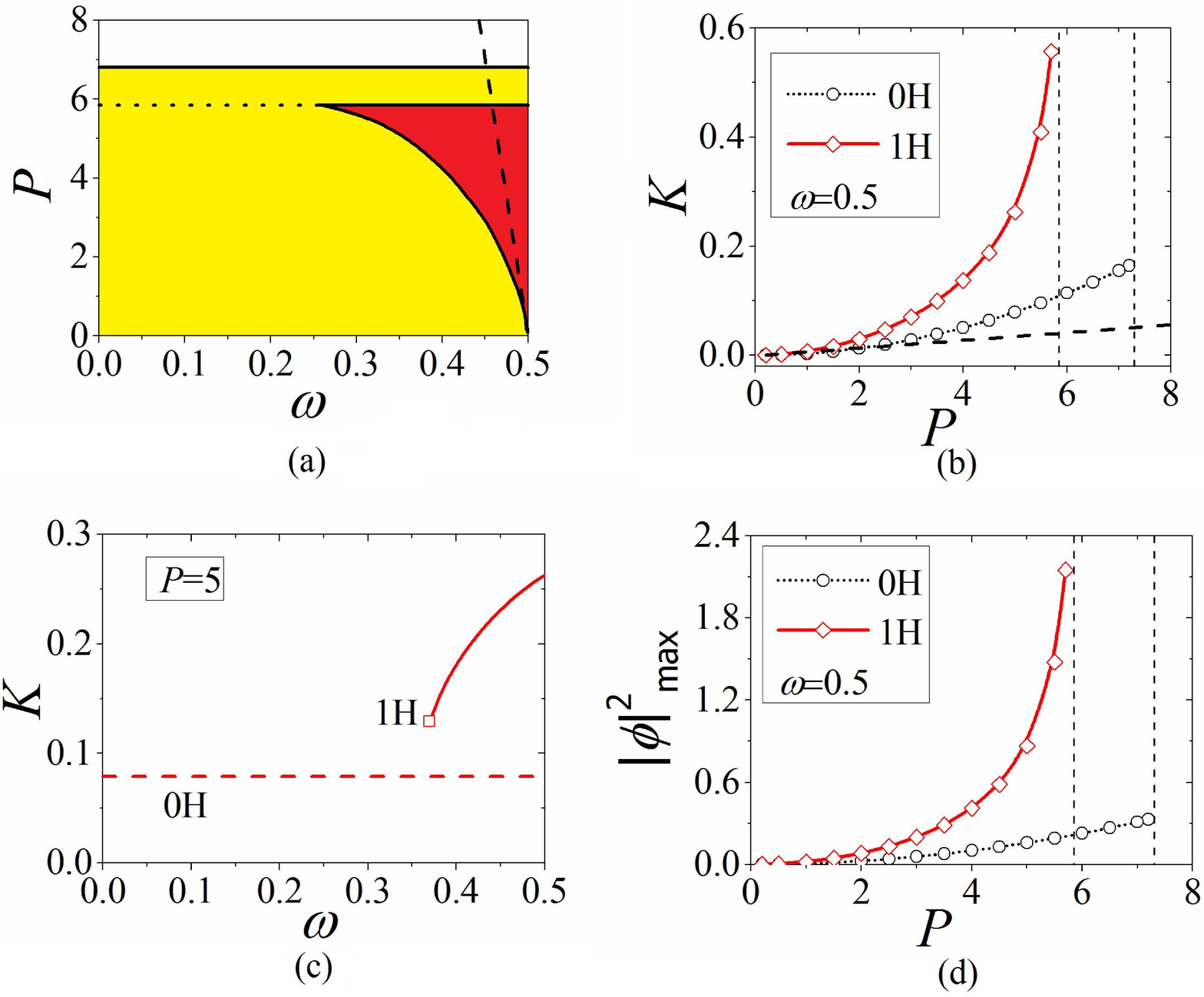}
\caption{(Color online) (a) Stability areas for 0H (quasi-1D temporal soliton ) and 1H solitary mode (quasi-2D bullet) in the plane of $(P,\omega)$. The yellow color area is the stable area solely for the 0H modes, and the red color area is the bistable area both for the 0H and 1H modes. No soliton solution is found in the white color area of the plane. The dashed line is $P=138.9(0.5-\omega)$, the slope of 138.9 is the slope of the border of the stability area of the 1H solitary mode at $(P,\omega)\rightarrow(0,0.5)$. The horizontal dotted line is the threshold norm of $P_{\mathrm{cr1}}=5.85$. (b) $K$ of 0H and 1H solitary modes versus $P$ at $\omega=0.5$. The dashed line is $K=0.0072P$, the slope of 0.0072 is the slope of the curves of $K(P)$ at $P\rightarrow0$. (c) $K$ of 0H and 1H solitary modes versus $\omega$ for $P=5$. (d) The peak value of the 0H and 1H modes versus $P$ at $\omega=0.5$. The two vertical dashed lines in these two panels are $P_{\mathrm{cr1}}=5.85$ and $P_{\mathrm{cr2}}=7.3$.}
\label{character1}
\end{figure}
\begin{figure}
\centering  \includegraphics[scale=0.5]{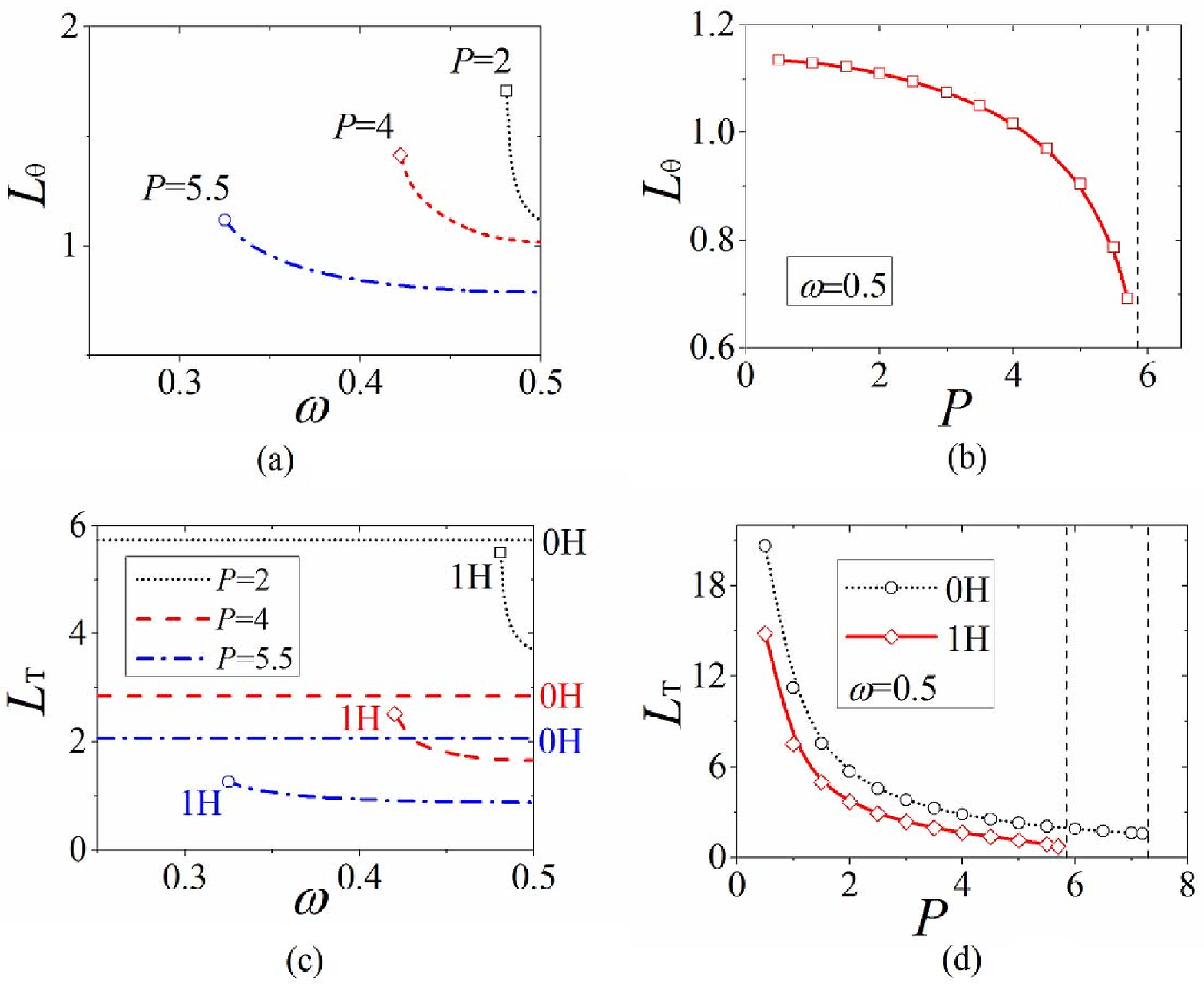}
\caption{(Color online) (a,b) $L_{\theta}$ of the 1H mode (quasi-2D bullet) versus $\omega$ and $P$, respectively. (c,d) $L_{T}$ for both solitary modes versus $\omega$ and $P$, respectively. }
\label{length}
\end{figure}

The solutions to stationary equation (\ref{NLS2}) were solved by the imaginary-time method \cite{ITP1,ITP2}. There are two types of solitary modes in the FRBZ. The first type is the stable 0H solitary mode, which can be found in the entire FRBZ. The input for this solitary mode is defined by the ansatz
\begin{eqnarray}
\phi_{0}=A\exp(-\alpha T^2), \label{input1}
\end{eqnarray}
where $A$ and $\alpha$ are positive real numbers. A typical example of the 2D intensity pattern of such a solitary mode with $(P,\omega)=(5.8,0.5)$ is shown Fig. \ref{example}(a). The amplitude of this type of solitary mode is shown by the blue solid line in Fig. \ref{example}(c). As expected, the 0H solitary mode has a uniform constant amplitude in the entire $\theta$ (spatial) domain, which makes its intensity variant only with $T$ and not with $\theta$. Therefore, solitary modes of this type can be considered quasi-1D objects and called quasi-1D temporal solitons.

The other type is the 1H solitary mode, which is found in the range of the right part of the FRBZ. The intensity of the 1H solitary mode varies both on $\theta$ and $T$, and the input for this solitary mode is
\begin{eqnarray}
\phi_{0}=A\exp(-\alpha T^2-\beta\theta^{2}),\label{input2}
\end{eqnarray}
where $\beta$ is also a positive real number. A typical example of such a solitary mode with $(P,\omega)=(5.8,0.5)$ is displayed in Fig. \ref{example}(b). One can see that the transverse width of the intensity pattern of the 1H solitary mode is much narrower than the domain of $\theta$. The amplitude of such solitary mode is shown in Fig. \ref{example}(c), in which the curves of the real and imaginary parts are even and odd in the domain of $\theta$, respectively. Figure \ref{example}(d) shows that the width of this 1H solitary mode is narrower than the width of the angular domain. Finally, figure \ref{example}(e) shows the 1H solitary mode has a narrower temporal width and much higher peak power than the 0H temporal mode does; the insets in this panel, which display the spectra of the eigenvalues of these two solitary modes, demonstrate that these two types of solitary modes are stable. Because the 1H solitary mode varies both in the spatial and temporal domains, it can be considered a quasi-2D object and termed a quasi-2D bullet.

Two stable solitary modes with the same magnitudes of $P$ and $\omega$ in Fig. \ref{example} imply that there should be a bistable area for these two modes in the parameter plane of $(P,\omega)$. Figure \ref{character1}(a) displays the stability area of these two types of solitary modes. As expected, the 1H mode bifurcates its stability area from the point of $(P,\omega)\equiv(0,0.5)$ and spreads towards left edge of the FRBZ with the increase of $P$. However, the extension of its stability area stops at $P=P_{\mathrm{cr1}}=5.85$, which is the well-known scaled threshold for \emph{Townes solitons} \cite{Townes1964}. The 1H modes, i.e., the quasi-2D bullets, collapse to a singular point when $P>P_{\mathrm{cr1}}$. The left edge of the stability area of the 1H mode is exactly at the point of $(P,\omega)=(P_{\mathrm{cr1}}, 0.25)$, which is the center of the FRBZ. The other type of solitary modes, i.e., the quasi-1D temporal soliton, features a different characteristic on its stability area. The 0H modes can stable in the entire FRBZ. Their collapses occur at $P>P_{\mathrm{cr2}}=7.3$, which is higher than the threshold of \emph{Townes solitons} \cite{Townes1964}. The difference between the $P_{\mathrm{cr1}}$ and $P_{\mathrm{cr2}}$ also demonstrates the truth that the 0H and 1H modes belong to the 1D and 2D schemes, respectively.

Figures \ref{character1}(b,c) display the $K$ of the 0H and 1H modes versus $P$ and $\omega$, respectively. The $K-P$ relationship in Fig. \ref{character1}(b) shows that these two types of solitary modes satisfy the Vakhitov-Kolokolov (VK) criterion \cite{VK,Xiangyu}, $dK/dP>0$, which is a well-known necessary condition for the stability of the soliton in the self-focusing nonlinear media. The $K-\omega$ relationship in Fig. \ref{character1}(c) shows that the $K$ of the 1H modes depend on the variable $\omega$, whereas the $K$ of 0H modes do not depend on $\omega$. The peak values of two types of solitary modes are also considered in Fig. \ref{character1}(d). The figure shows that the peak value of 1H mode is markedly higher than that of the 0H mode with an equal energy flow. Finally, it is also relevant to discuss the widths of two types of solitary modes in the domains of $\theta$ and $T$ for this special quasi-2D system. The width on $\theta$, i.e., transverse width, is defined as
\begin{eqnarray}
L_{\theta}=\left(\frac{\int_{-\pi}^{+\pi}\theta^2|u(\theta,0)|^2d\theta}{\int_{-\pi}^{+\pi}|u(\theta,0)|^2d\theta}\right)^{1/2},\label{Ltheta}
\end{eqnarray}
and width on $T$, i.e., the longitude width, is defined as
\begin{eqnarray}
L_{T}=\left(\frac{\int_{-T}^{+T}T^2|u(0,T)|^2dT}{\int_{-T}^{+T}|u(0,T)|^2dT}\right)^{1/2}. \label{LT}
\end{eqnarray}
Obviously, $L_{\theta}$ is only valid for the 1H modes because the 0H modes are homogeneous in the entire $\theta$ domain. While $L_{T}$ is valid for both solitary modes. Figure \ref{length} displays these two widths versus $\omega$ and $P$. Fig. \ref{length}(a) shows that the transverse width, $L_{\theta}$, of the 1H mode decreases with $\omega\rightarrow0.5$, and the minimum width is at $\omega=0.5$. Fig. \ref{length}(b) shows that $L_{\theta}$ decreases with the increase of $P$, which can be naturally understood by the action of self-focusing nonlinearity. Moreover, in Fig. \ref{length}(b), one can see that at $\omega=0.5$, the magnitude of $L_{\theta}$ of the 1H mode is far smaller than the transverse length of the $\theta$ domain $\sim2\pi$ for most values of $P$. This observation also indicates why the 1H mode is termed a quasi-2D object.
 Fig. \ref{length}(c,d) show the longitude width, $L_{T}$, for two types of solitary modes versus $\omega$ and $P$, respectively. For the 1H mode, one can see that the $L_{T}$ of the 1H modes decreases with both $\omega\rightarrow0.5$ and an increase in $P$. While for the 0H mode, their $L_{T}$ only depend on $P$ and do not depend on $\omega$. Therefore, one can conclude from Figs. \ref{character1} and \ref{length} that the 1H modes (i.e., the quasi-2D bullet) depend on both $T$ and $\omega$, whereas the 0H modes (i.e., quasi-1D temporal solitons) depend only on $T$. This conclusion is in accordance with their 2D and 1D properties.

The bistable area of these two types of solitary modes can be analyzed by adopting small-amplitude modes at $P\rightarrow0$ and $\omega\rightarrow0.5$ (i.e., the bottom right corner of the FRBZ). For convenience, we define $\delta=1/2-\omega\ll1/2$. The spatiotemporal wavefunction can be separated to $\phi(\theta,T)=w(T)U(\theta)$. When $P\rightarrow0$, $w(T)$ is a real and broad wavepacket, which can be assumed to be a constant, i.e., $w(T)\simeq c$ (where $c$ is a small enough real number), in the range of the soliton. Hence, the second derivative of $T$ in Eq. (\ref{NLS2}) can be neglected by $\partial^{2}\phi/\partial T^{2}=U\partial^{2}w/\partial T^{2}\simeq0$. According to the analysis in Sec. II, the corresponding stationary solutions of the transverse mode in the $\theta$ domain are sought as a combination of the 0H and 1H continuous-wave (CW) mode:
\begin{eqnarray}
U(\theta)=a_{0}+a_{1}e^{i\theta},\label{0H1HCW}
\end{eqnarray}
where $a_{0}$ is fixed to be real, whereas amplitude $a_{1}$ may be complex. Next, under the aforementioned conditions, equation (\ref{NLS2}) is separated into the following two algebraic equations while considering the balance of the 0H and 1H modes in the equation:
\begin{eqnarray}
&&K=c^{2}\left(a^{2}_{0}+2|a_{1}|^{2}\right),\label{0HCW}\\
&&K+\delta=c^{2}\left(2a^{2}_{0}+|a_{1}|^{2}\right). \label{1HCW}
\end{eqnarray}
Under this circumstance, the energy flow of the solitary modes is redefined as
\begin{eqnarray}
P=\int^{\infty}_{-\infty}w^{2}(T)dT\int^{\pi}_{-\pi}|u(\theta)|^{2}d\theta\simeq2\pi A(a^{2}_{0}+|a_{1}|^{2}), \label{0H1HPower}
\end{eqnarray}
where $A=\int^{\infty}_{-\infty}w^{2}(T)dT$. According to Eqs. (\ref{0HCW},\ref{1HCW}), the propagation constant
is related to the energy flow (\ref{0H1HPower}) of solitary mode,
\begin{eqnarray}
K={3\over4\pi(A/c^{2})}P-{\delta\over2},\label{analmu-P}
\end{eqnarray}
The solution of Eqs. (\ref{0HCW},\ref{1HCW}) with $a_{0}a_{1}\neq0$ give rise to the species of 1H mode with
\begin{eqnarray}
&&a^{2}_{0}={1\over3c^{2}}(2\delta+K)\nonumber\\
&&|a_{1}|^{2}={1\over3c^{2}}(-\delta+K). \label{anal2Dsoliton}
\end{eqnarray}
Equation (\ref{anal2Dsoliton}) predicts the border of the 1H mode at $|a_{1}|^{2}=0$ (i.e., at $\delta=K$). Thus, Eq. (\ref{analmu-P}) yields the location of this boundary in terms of the energy flow $P_{\min}=2\pi(A/c^{2})|\delta|$. Therefore,   Eq. (\ref{analmu-P}) predicts the coexistence of the 0H and 1H solitary modes, in the limit of rotation speeds $\delta\rightarrow0$ (i.e., $\omega\rightarrow0.5$) and $P\rightarrow0$, at
\begin{eqnarray}
P>P_{\min}=2\pi(A/c^{2})|\delta|. \label{analborder}
\end{eqnarray}
The dashed line in Fig. \ref{character1}(b) shows that, at $\delta=0$ (i.e., $\omega=0.5$), ${dK/ dP}|_{P\rightarrow0}\simeq0.0072$, which gives rise to $A/c^{2}\approx22.3$. This result predicts that the absolute value of the slope of the coexistence area at $(P,\omega)\equiv(0,0.5)$ is $\sim138.9$. This prediction is well in accordance with the numerical results in Fig. \ref{character1}(a) at $(P,\omega)\rightarrow(0,0.5)$ [see the dashed line in Fig. \ref{character1}(a)].

\begin{figure}
\centering  \includegraphics[scale=0.3]{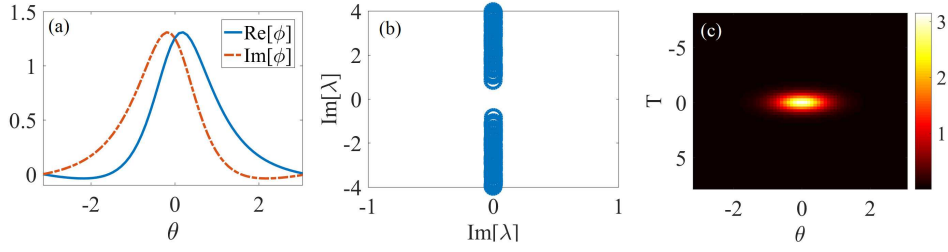}
\caption{(Color online) A typical example of the stable 1H solitary mode of the MM type for $(P,\omega)=(5.8, 0.5)$. (a) Cross-section of the amplitudes of the real (blue solid curve) and imaginary (red dash-dot curve) parts. (b) Spectra of the eigenvalue $\lambda$ of this solitary mode, which demonstrate its stability. (c) Intensity pattern of this solitary mode in the $(\theta, T)$ plane. }
\label{mixedmode}
\end{figure}

Because the 1H solitary modes are complex functions, the 1H solitary mode in this system can be applied to emulate the nonlinear dynamics of the solitary modes with spin-orbit coupling of the Rashba type in optics. Here, we rewrite the $U(\theta)=U_{\mathrm{R}}(\theta)+iU_{\mathrm{I}}(\theta)$, where $U_{\mathrm{R.I}}(\theta)$ are real functions, and substitute it into Eq. (\ref{NLS2}) to obtain a coupled equation for the real and imaginary parts:
\begin{eqnarray}
&&-{1\over2}{\partial^{2}\over\partial\theta^{2}}U_{\mathrm{R}} -\omega{\partial\over\partial\theta}U_{\mathrm{I}}-c^2 (U^{2}_{\mathrm{R}}+U^{2}_{\mathrm{I}})U_{\mathrm{R}}+K U_{\mathrm{R}}=0,\nonumber\\
&&-{1\over2}{\partial^{2}\over\partial\theta^{2}}U_{\mathrm{I}} +\omega{\partial\over\partial\theta}U_{\mathrm{R}}-c^2 (U^{2}_{\mathrm{R}}+U^{2}_{\mathrm{I}})U_{\mathrm{I}}+K U_{\mathrm{I}}=0. \label{SOCURI}
\end{eqnarray}
The type of the derivative couplings in Eq. (\ref{SOCURI}) are exactly in accordance with the Rashba spin-orbit coupling in 1D \cite{Yongping}. In fact, the amplitude of the 1H solitary mode in Fig. \ref{example}(c) (the real part denotes the fundamental component, and the imaginary part denotes the dipole component) can be termed a semi-dipole (SD) type solitary mode, which is supported by the 1D Rashba SO-coupling \cite{Rongxuan}. Their counterparts supported by the 2D Rashba SO-coupling are named semi-vortices, and they have attracted much interest in recent studies \cite{Xunda2016,yongyao20172,Bingjin2017}. The other type of solitary modes supported by the Rashba SO-coupling is named mixed modes (MMs). The 2D version of MMs is constructed by combining two vortical ingredients with $S_{+}=(0,-1)$ and $S_{-}=(0,+1)$ in the two components\cite{YongyaoNJP2017,Guihua20172}. In the 1D version, MMs may be obtained by inputting the ansatz of
\begin{eqnarray}
&&\phi^{\mathrm{R,I}}_{0}=\left[A_{1}\pm A_{2}\theta\exp(-\beta\theta^{2})\right]\exp(-\alpha T^2),\nonumber\\
&&\phi_{0}=\phi^{\mathrm{R}}_{0}+i\phi^{\mathrm{I}}_{0}(\theta,T),
\end{eqnarray}
where $A_{1,2}$ are real constants. A typical example of 1H solitary mode of the MM type is shown in Fig. \ref{mixedmode}. Numerical simulations demonstrate that the 1H modes of the SD type and MM type are complete degenerated. They have the same propagation constant, $K$, and the same intensity profile, $|u(\theta,T)|^{2}$, when their controlled parameters $P$ and $\omega$ are the same. Notice again that such spin-orbit coupling arose from a single component system, which is different from the previous works emulated from the two-component system by means of dispersive coupling or diffraction coupling \cite{Kartashov2015,Kartashov20152,SakaguichiNJP2016}.

It is relevant to estimate the actual parameters for this setting in the end of this section. If we adopt the wavelength of the optical field of 1.55 $\mu$m, the constant of group velocity dispersion is 40 ps$^{2}$/km, and the nonlinear parameter is 8 W/km \cite{Agrwal2001}. If we assume that $z=1$ is $\sim0.25$ m, one may obtain that $r_{0}=1$ is $\sim0.25$ mm, $T=1$ is $\sim0.1$ ps, the screw pitch of the twisted waveguide for $\omega=1$ is $\sim1.5$ m, and the unit of the $|\phi|^{2}$ is $\sim5$ mW. These parameters are realistic for a real setting.

\section{Collisions between the solitary modes}
\begin{figure}
\centering  \includegraphics[scale=0.3]{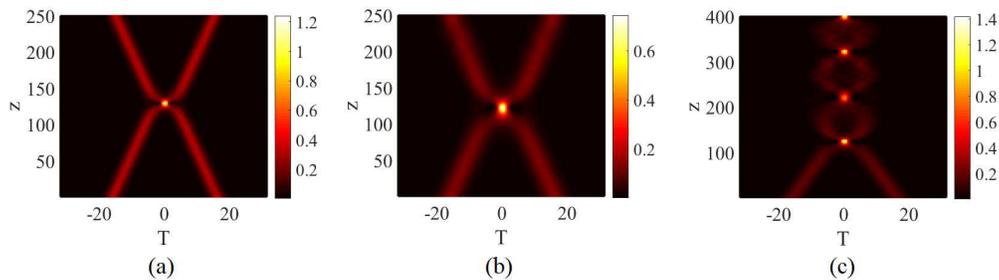}
\caption{(Color online) (a) Elastic collision of a pair of 0H solitary modes (quasi-1D temporal solitons) with $P=7$.  (b) Elastic collision of a pair of 1H solitary modes (quasi-2D bullets) with $P=2.5<P_{\mathrm{cr1}}/2$. (c) Inelastic collision (multi-collision) of a pair of 1H modes with $P\sim P_{\mathrm{cr1}}/2$. Here, in all panels, we fix $T_{0}=20$ and $\eta=0.1$. }
\label{collision}
\end{figure}

Collisions between these two types of solitary modes are conducted by assuming a pair of solitary modes with a separation $2T_{0}$ in the $T$ domain and an opposite direction kicks respectively onside them. In this case, the initial condition of the soliton is written as
\begin{eqnarray}
\psi(\theta, T,z=0)=\phi(\theta,T+T_{0})e^{i\eta T}+\phi(\theta,T-T_{0})e^{-i\eta T}, \label{initial}
\end{eqnarray}
where $\eta$ is the strength of the kick.

For the 0H modes, only elastic collisions are found when the power of the single solitary mode $P<P_{\mathrm{cr2}}$, Figure \ref{collision}(a) shows a typical example of such collision. For the 1H modes, elastic and inelastic are found when the energy flow of the single solitary mode smaller or close to $P_{\mathrm{cr1}}/2$. The inelastic collision is a multi-collision after the first collision. Typical examples of these two types of collisions are displayed in Fig. \ref{collision}(b,c). If the energy flow of the single 1H mode $P>P_{\mathrm{cr1}}/2$, then collapse occurs after the collision.

\section{conclusion}

The objective of this work was to study the spatiotemporal solitary modes in a twisted cylinder waveguide pipe with a self-focusing Kerr nonlinearity. The analysis was performed in the first rotational Brillouin
zone (FRBZ). We demonstrated that only two types of solitary modes, zero-harmonic (0H) and first-harmonic (1H), are found in the FRBZ. The 0H solitary mode, which is termed as a quasi-1D (one-dimensional) temporal soliton, can be found in the entire FRBZ when their energy flow is smaller than $P<7.3$. Their amplitudes are constant in the angular domain of the cylinder waveguide shell. Characteristics such as the propagation constant and the temporal width depend only on their energy flow and not on the rotating speed of the twisted waveguide.

The 1H solitary mode is termed a quasi-2D (two-dimensional) bullet. The spatial widths (in the angular domain) of these bullets are much narrower than the width of the entire angular domain. The characteristics of this type of solitary mode depend both on their energy flow and the rotating speed of the waveguide. These bullets are stable when their energy flow is smaller than $P=5.85$, the well-known scaled threshold norm of the \emph{Townes soliton}. The boundaries of the bistable area of these two types of solitary modes were also found in an analytical form by means of a two-mode approximation, in accordance with the numerical findings. By the analysis, we also demonstrate that the 1H solitary mode of this system can be applied to emulate the nonlinear dynamics of solitary modes with 1D Rashba spin-orbit coupling. Two types of degenerated states of the 1H solitary mode, semi-dipole and mixed mode, were revealed by the emulation.

Finally, collisions between the pair of 0H and 1H modes were discussed in the paper. The 0H mode pair features only the elastic collision, whereas the 1H mode pair could feature elastic and inelastic collisions when the total energy flows of the two modes smaller or close to the threshold norm of the \emph{Townes soliton}.

The present analysis can be extended in several directions.
First, a natural possibility is to consider the optical field from a single signal to a field from double signals. We can also study the two-component system with cross-phase modulation and linear mixing between them \cite{zhihuan2013}. Moreover, one can replace the linear mixing with dispersive coupling \cite{Kartashov2015,Kartashov20152} to emulate the 2D spin-orbit-coupling in the rotating system with a toroidal trapping.  Finally, one can add the Bragg-grating on the $z$-direction to the twisted waveguide shell and design some function settings \cite{shenhe2017,Zhigui2016,Hongji2015,Shenhe2013} for the spatiotemporal solitary mode against the twisted rate.

%
%

\end{document}